\documentclass{workshop}
\usepackage{rotating}
\usepackage{lscape}
\def\ga{\mathrel{\mathchoice {\vcenter{\offinterlineskip\halign{\hfil
$\displaystyle##$\hfil\cr>\cr\sim\cr}}}
{\vcenter{\offinterlineskip\halign{\hfil$\textstyle##$\hfil\cr
>\cr\sim\cr}}}
{\vcenter{\offinterlineskip\halign{\hfil$\scriptstyle##$\hfil\cr
>\cr\sim\cr}}}
{\vcenter{\offinterlineskip\halign{\hfil$\scriptscriptstyle##$\hfil\cr
>\cr\sim\cr}}}}}
\begin{document}

\title{Understanding superbursts \\ 
}

\author{
Jean in 't Zand
\\[12pt]  
%
SRON Netherlands Institute for Space Research, Sorbonnelaan 2, 3584 CA Utrecht, the Netherlands \\
%
\textit{E-mail: jeanz@sron.nl} 
}

\abst{Superbursts were discovered at the beginning of this
  millennium. Just like type-I X-ray bursts, they are thought to be
  due to thermonuclear shell flashes on neutron stars, only igniting
  much deeper. With respect to type-I bursts, they last 10$^3$ times
  longer, are 10$^3$ as rare, ignite 10$^3$ times deeper (in column
  depth) and are thought to be fueled by carbon instead of hydrogen
  and helium. Observationally, they are sometimes hard to distinguish
  from intermediate duration bursts which are due to pure helium
  flashes on cold neutron stars. So far, 26 superbursts have been
  detected from 15 neutron stars in low-mass X-ray binaries that also
  exhibit type-I bursts. They are very difficult to catch and only 2
  have been measured with highly sensitive
  instrumentation. Superbursts are sensitive probes of the neutron
  star crust and the accretion disk. The superburst phenomenon is not
  fully understood. Questions remain about the nature of the fuel, the
  collection of that fuel and the ignition conditions. The current
  state of affairs is reviewed and possible resolutions that lay ahead
  in the future discussed.}

\kword{Stars: neutron -- X-rays: bursts, binaries -- Nuclear reactions}

\maketitle
\thispagestyle{empty}

\section{Introduction}

Since 1969, 'type-I' X-ray bursts are being detected from space-borne
observatories between roughly 1 and 10 keV
\citep{belian1972,grindlay1976,matsuoka1980,makishima1981,gottwald1986,
  lewin1993,stroh2006,galloway2008}.  These bursts are due to
thermonuclear shell flashes of hydrogen and helium in the freshly
accreted upper layers of neutron stars in low-mass X-ray binaries
(LMXBs; \cite{hansen1975,lamblamb1978,wallace1981,fujimoto1981}), like
those on white dwarfs are responsible for classical novae. The
durations of most type-I bursts are between a few seconds and a few
minutes. The time profile of the X-ray emission is typically a fast
rise of duration a few seconds and an exponential-like decay. The
shell flash occurs at large ($\gg 1$) optical but small linear depth
(roughly 1 m) below the photosphere and temperatures may rise to order
1 GK. By the time the heat wave reaches the photosphere, the
temperature is a few tens of MK and the typical peak energy of the
photon spectrum is 5-10 keV. The flash itself lasts a fraction of
second, although it may take a few seconds to engulf the whole neutron
star. The burst decay phase essentially is the cooling of the neutron
star and actually does not follow an exponential function but a power
law (e.g., \cite{cumming2004,zand2014a}). The duration of the cooling
is primarily set by the amount of mass that is heated up or, in other
words, the column depth of the ignition (usually between 10$^8$ and
10$^9$~g~cm$^{-2}$).

\begin{figure}[t]
\centering
\includegraphics[height=\columnwidth,angle=270,trim=10mm 0mm 20mm 0cm]{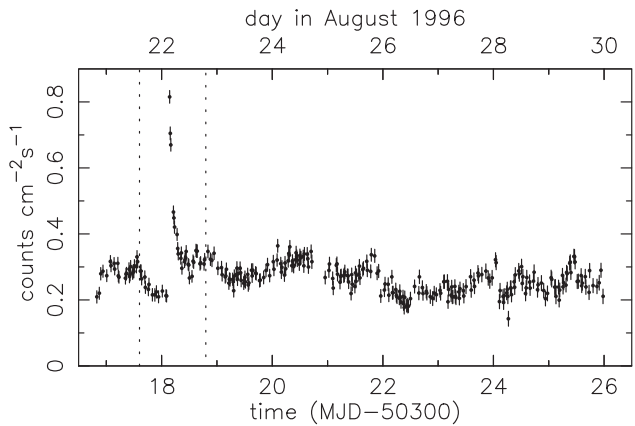}
\caption{Nine-day long light curve of 4U 1735-44 as observed with
  BeppoSAX-WFC, with the first detection of a superburst on August 22,
  1996 (from \cite{cor00})}
\label{fig1735}
\end{figure}

In the fall of 1996, the just launched two Wide Field Cameras (WFCs;
\cite{jager1997}) onboard the Italian-Dutch BeppoSAX observatory
\citep{boella1997} conducted a nine-day long observation of the
Galactic center region. Thanks to the $40\times40$ square degrees
field of view per camera, the WFCs could simultaneously observe about
half the Galactic low-mass X-ray binary population. One of the LMXBs
in the field of view was 4U 1735-44. Figure~\ref{fig1735} shows the
nine-day light curve resulting from that observation. It includes a
remarkable feature on August 22, 1996: a fast rise exponential-like
decay phenomenon with a duration of a few hours. \citet{cor00} found
this, and the spectrum, very reminiscent of a type-I burst except for
the duration which is $10^3$ as long. They proposed this as the
longest thermonuclear burst ever observed. This marks the discovery of
superbursts.

The discovery gave rise to searches in archival data from WFC and the
All-Sky Monitor (ASM) on RXTE, which yielded 12 more very long bursts
that were published between 2001 and 2004 (see Table~\ref{tab1}).
\citet{wijnands2001} introduced the term 'superburst'. \citet{cum01}
and \citet{stroh02a} introduced the first explanation of the
phenomenon as a thermonuclear shell flash fueled by carbon at a column
depth $\sim10^3$ times deeper than for type-I bursts.

In this paper, we briefly discuss the observational facts (\S
\ref{obs}) and theoretical considerations (\S \ref{the}) that define
our current understanding of superbursts, summarize the importance of
superburst research (\S \ref{why}) and touch on future prospects (\S
\ref{fut}).

%
%

\section{Observational overview}
\label{obs}

\subsection{Catalog \& recurrence time}

Table~\ref{tab1} presents a list of all 26 superbursts and their
characteristics that have been reported up to January 2017. The most
recent one is reported in these proceedings (\cite{iwakiri2016a}, see
also \cite{iwakiri2016b}). It is from 4U 1705-44, a LMXB that was
already predicted to be a superburster a decade ago
(\cite{zand03}). The 26 superbursts are emitted by 15 low-mass X-ray
binaries that are also emitters of type-I bursts. Currently, MAXI on
the ISS is the most efficient superburst discovering machine. All six
superbursts since 2011 were discovered with this device.

Five sources have exhibited multiple superbursts. They have high
accretion rates except 4U 0614+09. The range of superburst recurrence
times is between 10 d (for GX 17+2) to 10.5 yr (for 4U 1820-30). The
average recurrence time is 4 yr, but this should be taken as an upper
limit because there are data gaps. \citet{zand03} perform a
statistical analysis of the recurrence time and find $2\pm1$
yr. \citet{keek2006} determine on a source-by-source basis and on the
basis of the BeppoSAX-WFC database a lower limit to the recurrence
time of usually 2 months. These determinations can probably be
improved upon considerably with the much larger data sets that is now
available through for instance INTEGRAL and MAXI.

\begin{sidewaystable*}
\caption{Catalog of 26 superbursts from 15 sources, detected up to
  January 2017. \label{tab1}}
\begin{center}
\small
\vspace{-3mm}
\begin{tabular}{llllllllllll} \hline\hline\\[-6pt]
Source      & Tran- & $P_{orb}$ & Date (Instr.)    & Onset & Decay     & $L_{peak}$ & $\dot{M}$  & Nearest   & $y$ & $E$ & Ref.$^\ddag$ \\
            & sient? &(min)    &                       & obs.? & Time      & ($10^{38}$   & (edd.)     & burst (d) & (10$^{12}$ & (10$^{17}$  & \\
            &       &         &                       &       & (hr)      & erg~s$^{-1}$)&            &           & g~cm$^{-2}$) & erg g$^{-1}$) & \\
\hline
4U 0614+09  &   & 51?          & 2005-03-12(ASM)      &       & 2.1    & $>0.1$      & $<0.01$    & -367/+19 & 0.2 & 5   & 1,a   \\
            &   &              & 2014-11-03(MAXI)     &       & 5.2    & 2.8         &            &          &     &     & 2     \\
4U 1254-69  &   & 236          & 1999-01-09(WFC)      & y     & 6.0    & 0.4         & 0.13       & -51/+125 & 2.7 & 1.5 & 3, 4, b        \\
4U 1608-52  & y &              & 2005-05-05(ASM+HETE) & y     & 4.5    & 0.5         & 0.03       & -57/+104 & 2.8 & 1.6 & 5     \\
4U 1636-53  &   & 228          & 1996-06-19(ASM)      &       & 3.1    & 1.0         &            &   +96    &     &     & 6, c     \\
            &   &              & 1997-07-13(ASM)      &       & 1.8    & $>0.9$      &            & -122/+68 &     &     & 7     \\
            &   &              & 1999-05-26(ASM)      &       & 2      & 0.8         &            &  -27/+15 &     &     & 8     \\
            &   &              & 2001-02-22(PCA+ASM)  & y     & 1.5    & 0.7         &            & -2/+23   & 0.48& 2.6 & 9, 4         \\
4U 1705-44  &   &              & 2016-10-22(MAXI)     &       & 2.2    & $>1.0$      &            &          &     &     & 10     \\
KS 1731-260 &$\sim$&           & 1996-09-23(WFC)      & y     & 2.7    & 1.4         & 0.1        & -6/+34   & 1.0 & 1.9 & 11, 4       \\
4U 1735-44  &   & 279          & 1996-08-22(WFC)      &       & 1.4    & 1.5         & 0.25       & ../+374  & 1.3 & 2.6 & 12, 4, d       \\
GX 3+1      &   &              & 1998-06-09(ASM)      &       & 1.6    & 0.8         & 0.2        & -62/+94  &     &     & 13 \\
GX 17+2     &   & 10d?         & 1996-09-14(WFC)      &       & 1.9    & 1.0         & 1          &      +2  &     &     & 14, e \\
            &   &              & 1999-09-23(WFC)      &       & 1.0    & 1.3         & 1          &     +10  & 0.6 & 1.8 & 14 \\
            &   &              & 1999-10-01(WFC)      & y     & 0.7    & 1.7         & 1          &      +2  &     &     & 14, 4 \\
            &   &              & 2000-09-08(WFC)      & y     & 2.2    & 1.8         & 1          &     +12  &     &     & 14, 4 \\
EXO 1745-248& y &              & 2011-10-24(MAXI+BAT) &       & 10     & 0.7         & $<0.01$    &          & 1.0 & $>1$& 15, 16 \\
SAX J1747.0-2853&y&            & 2011-02-13(JEMX+MAXI)& y     & 4.2    & 3           & 0.1        & -711/+25 &     &     & 17, 2 \\
4U 1820-30  &   & 11           & 1999-09-09(PCA)      & y     & 1      & 3.4         & 0.1        & -168/+167& 1   & 10  & 18, f \\
            &   &              & 2010-03-17(MAXI+ASM)$^\ast$ & & 0.5    & $>3.3$      & 0.15       &  +1549  &     &     & 19 \\
Ser X-1     &   &              & 1997-02-28(WFC)      &       & 1.2    & 1.6         & 0.2        & -162/+34 & 0.55& 2.3 & 20, 4 \\
            &   &              & 1999-08-09(ASM)      &       & 3.6    & $>0.19$     & 0.15       &   +309   &     &     & 7 \\
            &   &              & 2008-10-14(ASM)      &       & 1.4    & $>0.26$     & 0.13       &   +55    &     &     & 7 \\
            &   &              & 2011-12-06(MAXI)     &       & 2.3    & 0.9         & 0.21       &          & 2.1 & 4   & 10 \\
SAX J1828.5-1037&?&            & 2011-11-12(MAXI)     &       & 2.3    & 0.7         & $<0.01$    &          &     &     & 21 \\
Aql X-1     & y & 1137         & 2013-07-20(MAXI)     &       & 4.3    & 1.0         & 0.1        &   +389   &     &     & 2, g \\
\hline
\end{tabular}
\end{center}
\small
\renewcommand{\baselinestretch}{1.}
$^\ast$According to \citet{serino2016}, this is not a superburst. It
is listed here, because we believe that the alternative, it being an
intermediate duration burst, is less likely given the high accretion
rate. $^\ddag$References (numeric for superburst data,
alphabetical for orbital period): 1 - \citet{kuu10}, 2- \citet{serino2016}, 3
- \citet{zand03}, 4 - \citet{cumming2004}, 5 - \citet{keek2008}, 6 -
\citet{wijnands2001}, 7 - \citet{kuu09b}, 8 - \citet{kuulkers2004}, 9 -
\citet{stroh02b}, 10 - \citet{iwakiri2016a}, 11 - \citet{kuu02a}, 12 -
\citet{cor00}, 13 - \citet{kuu02c}, 14 - \citet{zand2004}, 15 -
\citet{altamirano2012}, 16 - \citet{serino2012}, 17 - \citet{che11},
18 - \citet{stroh02a}, 19 - \citet{zand11}, 20 - \citet{cor02}, 21 -
\citet{asada2011}, a - \citet{shabaz2008}, b - \citet{motch1987}, 
c - \citet{jvp90}, d - \citet{corbet1986}, e - \citet{reba2002},
f - \citet{pried1986}, g - \citet{chevalier1991}
\normalsize
\end{sidewaystable*}

\subsection{Host binaries}
\label{hostbinaries}

While in the early years (2000-2004) host binaries of superbursts were
all found to be LMXBs that are persistently accreting at a level of at
least 0.1 times the Eddington limit, the picture changed as the data
grew. There are now 4 transients among the 15 superbursters, and one
system (SAX J1828.5-1037) with an unknown nature, but for certain with
a low long-term average accretion rate like the 4 transients. Of the 7
host binaries with (tentatively) known orbital periods, 2 are
ultracompact X-ray binaries (UCXBs), meaning that the composition of
the donors, and therefore that on on the neutron star, is strongly
deficient in hydrogen (4U 1820-30 and 4U 0614+09).

There are a number of prolific bursters that are semi-persistent and
did not exhibit a superburst yet: 4U 1728-34, EXO 0748-676 (off since
2011), 4U 1702-429, 1E 1724.3045 (in Terzan 2), A 1742-294, 4U
1812-12, GS 1826-24 and Cyg X-2. While it is too early to derive a
physical meaning of this, it is something to keep in mind. It may be
related to the question of fuel accumulation for superbursts and
ignition conditions. \citet{zand03} did an investigation of the
average $\alpha$ parameter among a number of persistent
bursters. $\alpha$ is the ratio of the gravitational energy released
by accretion since the last burst to the nuclear energy released
through the present burst.  It should be between about 30 and 200
\citep{lewin1993}. \citet{zand03} found a clear distinction between
superbursters and non-superbursters.  The former ones have a
significantly higher $\alpha$ value ($\ga 1000$).

\subsection{Distinguishing superbursts from other long bursts}

\begin{figure}[t]
\centering
\includegraphics[width=\columnwidth,angle=0,trim=0mm 1mm 0mm 0mm,clip]{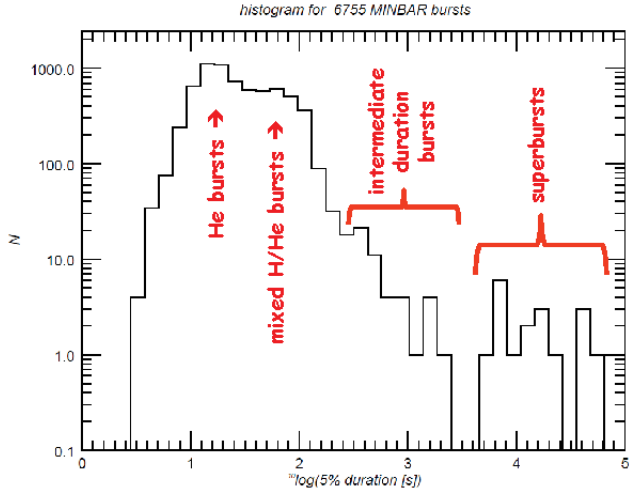}
\vspace{-3mm}
\caption{Preliminary histogram of burst durations as determined from
  bursts in the MINBAR archive (Galloway et al., in prep.)}
\label{figdur}
\end{figure}

After the discovery of superbursts, another kind of thermonuclear
X-ray burst was discovered that is also long but generally not as long
as superbursts: intermediate duration bursts \citep{zand05,cum06}, the
qualification {\em intermediate} referring to a duration between that
of type-I bursts and superbursts (see Fig.~\ref{figdur}). Since this
may incur uncertainty in the identification of short superbursts, we
discuss this somewhat more.

Most intermediate duration bursts are thought to result from the
ignition of thick helium piles on cold neutron stars. Due to the low
temperature, ignition is reached at higher pressure and larger column
depths, and the helium needs much more time to reach ignition (days to
weeks instead of hours) and bursts last longer. The thicker piles and
the fast $3\alpha$ helium-burning nuclear reaction will usually result
in very high nuclear powers that easily surpass the Eddington limit
and result in very strong photospheric expansion. The low temperatures
go hand in hand with low accretion rates. All these circumstances are
found in ultracompact X-ray binaries with orbital periods shorter than
about one and a half hour \citep{nrj86} and that is indeed what
observations show \citep{zand2007,zand10}.

\citet{peng2007} predict a small regime in accretion rate of H-rich
systems ($\approx0.003$ times Eddington) where also long helium bursts
can happen after long series of pure H bursts. However, this is a very
small range of allowed accretion rates and pure H-bursts have never
been detected yet. Nevertheless, there are sporadic reports of
intermediate duration bursts from H-rich systems
\citep{degenaar2010,chenevez2007}.

Identifying superbursts, particularly when data coverage is sporadic,
can be cumbersome.  One should, for instance, be careful about long
bursts that are discovered from UCXBs at low mass accretion
rates. These may be intermediate duration bursts. In fact, in a few
cases superburst detections had to be revised, see the careful
evaluation by \citet{serino2016} . Furthermore, in my opinion this
qualifies the identifications as superbursts in 4U 0614+09 and SAX
J1818-1036 as less certain (see also \cite{kuu10} for 4U 0614+09).

\subsection{Peak luminosities}

Roughly 20\% of type-I bursts have peak luminosities near the
Eddington limit \citep{galloway2008}: $2.0\times10^{38}$ erg~s$^{-1}$
for H-rich atmospheres and 3.4$\times10^{38}$~erg~s$^{-1}$ for H-poor
atmospheres. The situation is different for superbursts. All
superbursts except the PCA one from 4U 1820-30 \citep{stroh02a} and
possible the JEM-X burst of SAX J1747.0-2853 \citep{che11} are
sub-Eddington. This immediately shows that the fuel layer is not
burning completely.

\subsection{Precursors}

Most superbursts are discovered with low duty-cycle instruments,
particularly the ASM on RXTE and MAXI on the ISS. These devices
observe more than 80\% of the sky every 90-min satellite orbit, but
only for about 1 min. Therefore, it is easy to detect superbursts
because they generally last longer than 90-min, but it is also
difficult to catch the onset of superbursts. The onset has been
observed in 8 of the 26 superbursts. Interestingly, in each of these
cases the onset is marked by a short burst. This is often called a
precursor, but actually there is only one case (the PCA superburst of
4U 1820-30) where there is truly a brief period without burst emission
between the precursor and the main burst.

\begin{figure}[t]
\centering
\includegraphics[height=\columnwidth,angle=270,trim=10mm 0mm 14mm 0cm]{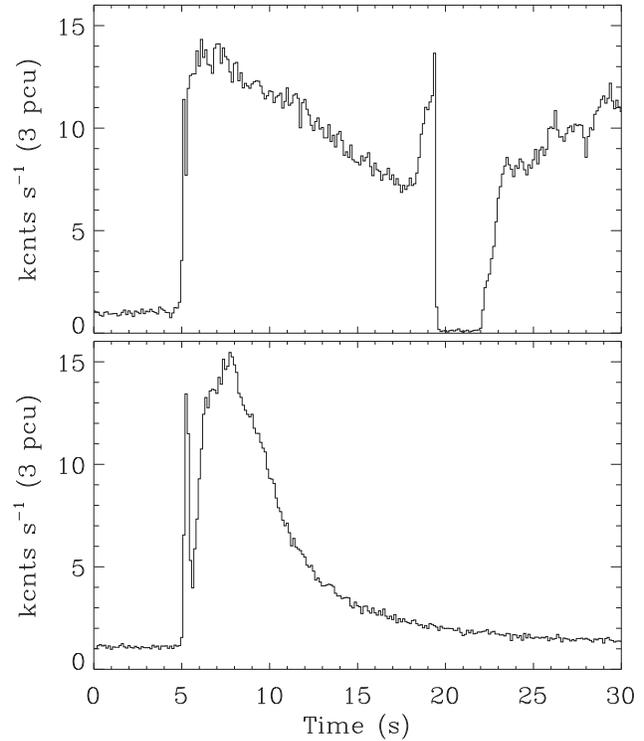}
\vspace{-8mm}
\caption{PCA-measured light curves of two bursts from 4U 1820-30. Top
  panel: first 30 s of the superburst. Bottom panel: a type-I
  burst from the same source. From \citet{stroh02a}.}
\label{figprecursors}
\end{figure}

Figure~\ref{figprecursors} shows the onsets of a superburst and an
ordinary burst from 4U 1820-30, detected with the high-throughput
PCA. The superburst onset is characterized by the precursor (from 5-18
s) and the superburst (starting at 18 s), and the dips in both these
bursts. The dip in the precursor is very short (less than the time
resolution of $\frac{1}{8}$ s) and possibly not complete to the
pre-burst flux level. The dip in the superburst drops to below the
pre-burst level. Both of these dips are consistent with photospheric
expansion \citep{lewin1984} with adiabatic cooling, whereby the
cooling in the second dip is so strong that the X-ray signal is lost
\citep{keek2012b}. The drop to below the pre-burst level is due to the
photosphere covering up the X-ray emitting part of the accretion disk
(e.g., \cite{zand10}).

Alternatively, \citet{weinberg2007} attribute the first brief dip as a
pause between a spike due to a shock breakout and a prematurely
ignited type-I pure-helium burst. However, \citet{keek2011} found that
the heat released by the fallback of the photosphere after the shock
breakout may be sufficient to ignite the type-I burst and that this
occurs on time scales much shorter (10$^{-5}$~s) than the dip time
scale. After analyzing the spectral evolution during the dip,
\citet{keek2012b} concluded that it could be attributed to
photospheric expansion.  \citet{keek2012b} also found that the
precursor is more energetic than ordinary type-I bursts and,
therefore, cannot be powered solely by the burning of accreted
helium. They attribute the additional energy to the shock heating,
supporting numerical models that predict that superbursts result in a
detonation \citep{weinberg2006b} and shock that generates enough heat
or overpressure to power the precursor.  A similar onset is observed
in the other superburst detected with the PCA.

\subsection{Burst quenching}

Some superbursters are prolific emitters of type-I bursts at the time
of the superburst, with recurrence times of merely a few hours. These
include KS 1731-26 and 4U 1636-536. But it is noticeable that this
emission of type-I bursts is quenched for a considerable period of
time after the superburst \citep{kuu02a,cor02}, namely about one
month. Apparently, the superburst influences the nuclear burning for a
considerably longer time than when its emission is visible.

Measuring the duration of burst quenching is difficult. One needs a
substantially high duty cycle to detect type-I bursts. This is not
easily accessible with regular all-sky monitors. Quench time
measurements are very interesting, though. They provide a clean means
to observe the transition between stable and unstable nuclear burning
(see \cite{kee12a}).

\section{Theory}
\label{the}

As with type-I bursts, the general picture is clear of how superbursts
come about (a deep thermonuclear shell flash), but there are two
essential issues that need resolution.

\subsection{Inferring basic physics parameters}

\begin{figure}[t]
\centering
\includegraphics[height=\columnwidth,angle=270,trim=0mm 15mm 15mm 24mm]{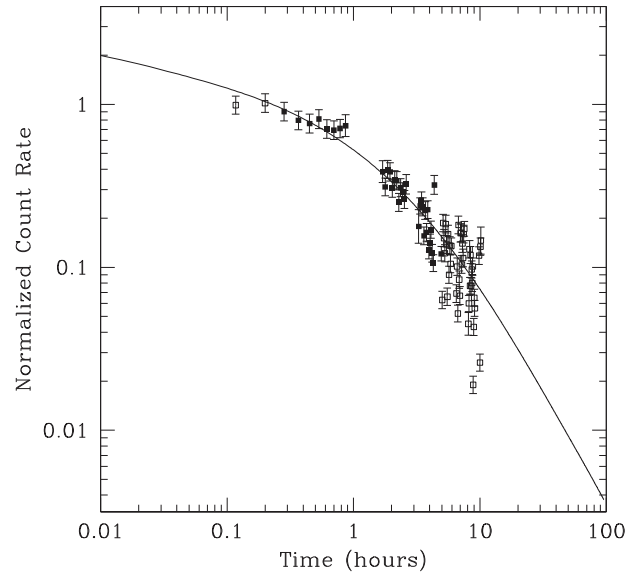}
\caption{The decay of the superburst from 4U 1735-44 (see
  Fig.~\ref{fig1735} fitted with a model parametrized with $y_{12}$
  and $E_{17}$ (see text and Table~\ref{tab1}). From \citet{cum06}.}
\label{figdecay}
\end{figure}

\citet{cumming2004} and \cite{cum06} showed that it is possible to
infer from the light curve, in particular the peak luminosity and the
decay profile, the ignition column depth
$y_{12}=y/10^{12}$~g~cm$^{-2}$ and energy yield
$E_{17}=E/10^{17}$~erg~g$^{-1}$ of the heated matter. The higher the
energy yield is, the higher the peak luminosity (up to the Eddington
limit).  The deeper the ignition is, the longer the duration. An
example of a fitted superburst decay is shown in
Fig.~\ref{figdecay}. Table~\ref{tab1} shows the values of these two
parameters for many superbursts. Due to the lack of onset coverage, it
is often difficult to obtain reasonable constraints on $y_{12}$ and
$E_{17}$, particularly the latter. Therefore, these numbers lack in
many cases.

\citet{keek2015} improved the light curve diagnostic power by
including the slope of the temperature relation with depth as a free
parameter. This impacts in particular the rise phase of the light
curve.

The expected values for CNO burning and $3\alpha$ process are
$E_{17}=64$ and 15, respectively. This is 1 to 2 orders of magnitude
larger than the observed values for superbursts. The energy yield for
nuclear burning of carbon to iron-group elements is about 10. The
superburst from 4U 1820-30 has the very same value. 

\subsection{What fuel is burning?}

The longer duration points to a deeper ignition, a higher density and
a higher temperature than for type-I bursts. This makes it easier to
overcome the Coulomb barrier of heavier elements than hydrogen and
helium. The next most abundant element, certainly after the burning of
hydrogen to helium through the CNO cycle and the burning of helium
through the $3\alpha$ process, seems carbon and this is the fuel that
was considered by \citet{stroh02a} and \citet{cum01}.

The $E_{17}$ values are more or less consistent with carbon burning.
They range between 1.5 and 10. This implies mass fractions of burnt
carbon between 15 and 100\%. Part of the remaining composition are the
heavy isotopes needed for the right ignition depth \citep{cum01} and
whose photo-disintegration, furthermore, might contribute up to half
the superburst energy, possibly lowering the required carbon in the
fuel \citep{schatz2003}.

\subsection{Issue 1: how to obtain and maintain enough fuel?}

If temperature becomes too high, the Coulomb barrier of carbon atoms
is easier overcome by ambient protons and alpha particles through the
reactions $^{12}$C(p,$\gamma$)$^{13}$N and
$^{12}$C($\alpha$,$\gamma$)$^{16}$O, thus destroying the carbon. This
can easily happen during helium flashes since the carbon and helium
are in the same layer. Thus, helium flashes are responsible for both
the production and destruction of carbon. Generally, it is thought
that the only manner in which carbon destruction can be prevented is
by preventing temperature to grow too large during $3\alpha$
burning. This is only possible during stable helium burning. There is
observational evidence for that, through the measurements of the burst
$\alpha$ parameter (see \S \ref{hostbinaries}). Also theoretically,
improvements are made in the understanding. \citet{stevens2014}
proposes that the rp-process consumes all protons before they get a
chance to capture on carbon.  \citet{keek2016} introduces a new regime
of stable hydrogen burning that increases the temperature in the
H-depleted layer underneath somewhat, yielding higher $3\alpha$ rates
without going runaway.

\subsection{Issue 2: how to ignite the fuel?}

For diluted carbon mixtures, as suggested by the $E_{17}$ values, it
is difficult to reach ignition conditions. The temperature at a depth
of $y_{12}\approx1$ is simply not high enough to overcome the Coulomb
barrier. This temperature is primarily set by a heat flow from the
crust, where pycnonuclear reactions and electron captures provide a
heat source of $Q_b=0.1-0.6$~MeV/nucleon. The power scales with the
accretion rate. \citet{cum01} initially resolved this issue by
proposing reduced conductivity of the ignition layer, thus reducing
the cooling of the layer. Then the crustal heating may be enough to
reach ignition temperatures, provided the mass accretion rate is in
excess of 0.1 times Eddington.

However, two recent findings aggravate the issue again. First, the
detection of superbursts from transient accretors, starting with 4U
1608-52 \citep{keek2008} and with an extreme case of EXO 1745-248
\citep{serino2012,altamirano2012}. The accretion rate, averaged over
the superburst recurrence time, is (much) lower for transients so that
crustal heating is accordingly weaker. Second, recent calculations of
neutrino cooling through the URCA process in the crust
\citep{schatz2014} and the deep ocean \citep{deibel2016} show it to be
much more efficient than previously thought.

The solution may come from finding shallow heating processes, for
instance due to rotational mixing and thus deeper CNO burning
\citep{keek2009} or freeze out of heavy elements at the bottom of the
ocean that induces convection which heats up the superburst layer
\citep{medin2011}, but as yet the issue of unreachable carbon ignition
conditions remains.

\section{Why study superbursts?}
\label{why}

As may be clear from above, studying superbursts is interesting in its
own right, but it is also very useful for addressing a wide variety of
scientific questions, for instance: 1) They are related to the same
nuclear process (explosive carbon burning) that is thought to be
responsible for type Ia supernovae. Understanding the ignition will
improve our understanding of type Ia SN ignition; 2) Since the
ignition is close to the neutron star crust, superburst
characteristics depend on the thermal properties of the outer crust
and, thus, provide a diagnostic of that crust (e.g., \cite{cum06}); 3)
Superbursts are sensitive probes of the neutron star spin and binary
orbit, through the detection of transient ms oscillations during
superbursts \citep{stroh02b}; 4) Superbursts can be used as seismology
probes of the neutron star interior, through the detection of
oscillations other than due to the spin \citep{stroh14}; 5)
Superbursts can be used as probes of the accretion disk. They
irradiate and heat up the accretion disk which can be observed through
reflection features in the superburst spectrum (e.g.,
\cite{ballantyne2004,keek2014}). This irradiation happens by a simple
spectral shape (black body) and over a broad range of tractable
temperatures; 6) The cooling of the neutron star envelope after a
superburst has an observationally convenient time scale (days) to
relatively easy probe ignition conditions of hydrogen and helium
burning \citep{kee12a}.

\section{Future}
\label{fut}

Only 11 superbursts have an observational coverage that is better than
10\%, strongly affecting measurements of peak luminosities, quench
times and onset profiles. Only 2 superbursts have been observed with
an effective area of significantly more than $\sim$100 cm$^2$,
strongly affecting sensitive measurements of the spectrum (e.g.,
\cite{keek2014}) and interesting variability such as burst
oscillations \citep{stroh02b}. Although MAXI and other instruments are
very useful in constraining superburst recurrence times and ignition
depths, it is obvious that much is to be gained from measurements with
higher duty cycles and larger sensitivity. This may be obtained
through 1) all-sky monitors that have large sky coverage and
reasonable effective area $\ga 100$~cm$^2$ in the 1-10 keV band and 2)
large X-ray telescopes with quick read-out times that spend
substantial amounts of observing time on the population of potential
superbursters (bursters with high $\alpha$ values) or can be quickly
brought on target after the onset of a superburst. One concept
platform where both of these types of instruments/observations are
foreseen in a optimum manner is LOFT (e.g.,
\cite{feroci2016,zand2015}), with spin off concepts eXTP
\citep{zhang2016} and Strobe-X \citep{wilson2017}. In the mean time,
Astrosat \citep{agr06} and NICER \citep{gendreau2012} are and will be
valuable assets in the study of superbursts if caught (see also
\cite{keek2016b}).

\vspace{3mm}\noindent {\em Acknowledgements.} Laurens Keek and Motoko
Serino are thanked for useful comments on an early draft of this
paper.  This paper uses preliminary analysis results from the Multi-
INstrument Burst ARchive (MINBAR), which is supported under the
Australian Academy of Science’s Scientific Visits to Europe program,
and the Australian Research Council’s Discovery Projects and Future
Fellowship funding schemes.


\begin{thebibliography}{69}
\expandafter\ifx\csname natexlab\endcsname\relax\def\natexlab#1{#1}\fi

\bibitem[{{Agrawal}(2006)}]{agr06}
{Agrawal}, P.~C. 2006, AdSR, 38, 2989

\bibitem[{{Altamirano} {et~al.}(2012){Altamirano}, {Keek}, {Cumming},
  {Sivakoff}, {Heinke}, {Wijnands}, {Degenaar}, {Homan}, \&
  {Pooley}}]{altamirano2012}
{Altamirano}, D., {et~al.} 2012, MNRAS, 426, 927

\bibitem[{{Asada} {et~al.}(2011){Asada}, {Negoro}, {Sugizaki}, {Matsuoka},
  {Mihara}, {Serino}, {Nakahira}, {Yamamoto}, {Sootome}, {Ueno}, {Tomida},
  {Kohama}, {Ishikawa}, {Kawai}, {Morii}, {Sugimori}, {Usui}, {Toizumi},
  {Aoki}, {Song}, {Yoshida}, {Yamaoka}, {Tsunemi}, {Kimura}, {Kitayama},
  {Nakajima}, {Suwa}, {Sakakibara}, {Ueda}, {Hiroi}, {Shidatsu}, {Tsuboi},
  {Matsumura}, {Yamauchi}, {Nishimura}, \& {Hanayama}}]{asada2011}
{Asada}, M., {et~al.} 2011, ATel, 3760

\bibitem[{{Ballantyne} \& {Strohmayer}(2004)}]{ballantyne2004}
{Ballantyne}, D.~R., Strohmayer, S.~E. 2004, ApJL, 602, L105

\bibitem[{{Belian} {et~al.}(1972){Belian}, {Conner}, \& {Evans}}]{belian1972}
{Belian}, R.~D., {et~al.} 1972, ApJL, 171, L87

\bibitem[{{Boella} {et~al.}(1997){Boella}, {Butler}, {Perola}, {Piro},
  {Scarsi}, \& {Bleeker}}]{boella1997}
{Boella}, G., {et~al.} 1997, A\&As, 122

\bibitem[{{Chenevez} {et~al.}(2011){Chenevez}, {Brandt}, {Kuulkers},
  {Alfonso-Garzon}, {Beckmann}, {Bird}, {Courvoisier}, {Del Santo}, {Domingo},
  {Ebisawa}, {Jonker}, {Kretschmar}, {Markwardt}, {Oosterbroek}, {Paizis},
  {Pottschmidt}, {Sanchez-Fernandez}, \& {Wijnands}}]{che11}
{Chenevez}, J., {et~al.} 2011, ATel, 3183

\bibitem[{{Chenevez} {et~al.}(2007){Chenevez}, {Falanga}, {Kuulkers}, {Walter},
  {Bildsten}, {Brandt}, {Lund}, {Oosterbroek}, \& {Zurita
  Heras}}]{chenevez2007}
{Chenevez}, J., {et~al.} 2007, A\&A, 469, L27

\bibitem[{{Chevalier} \& {Ilovaisky}(1991)}]{chevalier1991}
Chevalier, C., Ilovaisky, S.A. 1991, A\&A, 251, L11

\bibitem[{{Corbet} {et~al.}(1986)}]{corbet1986}
Corbet, R.H.D., et al. 1986, 222, 15P

\bibitem[{{Cornelisse} {et~al.}(2000){Cornelisse}, {Heise}, {Kuulkers},
  {Verbunt}, \& {in 't Zand}}]{cor00}
{Cornelisse}, R., {et~al.} 2000, A\&A, 357, L21

\bibitem[{{Cornelisse} {et~al.}(2002){Cornelisse}, {Kuulkers}, {in 't Zand},
  {Verbunt}, \& {Heise}}]{cor02}
{Cornelisse}, R., {et~al.} 2002, A\&A, 382, 174

\bibitem[{{Cumming} \& {Bildsten}(2001)}]{cum01}
{Cumming}, A., Bildsten, L. 2001, ApJL, 559, L127

\bibitem[{{Cumming} \& {Macbeth}(2004)}]{cumming2004}
{Cumming}, A., Macbeth, J. 2004, ApJL, 603, L37

\bibitem[{{Cumming} {et~al.}(2006){Cumming}, {Macbeth}, {in 't Zand}, \&
  {Page}}]{cum06}
{Cumming}, A., {et~al.} 2006, ApJ, 646, 429

\bibitem[{{Degenaar} {et~al.}(2010){Degenaar}, {Jonker}, {Torres}, {Kaur},
  {Rea}, {Israel}, {Patruno}, {Trap}, {Cackett}, {D'Avanzo}, {Lo Curto},
  {Novara}, {Krimm}, {Holland}, {de Luca}, {Esposito}, \&
  {Wijnands}}]{degenaar2010}
{Degenaar}, N., {et~al.} 2010, MNRAS, 404, 1591

\bibitem[{{Deibel} {et~al.}(2016){Deibel}, {Meisel}, {Schatz}, {Brown}, \&
  {Cumming}}]{deibel2016}
{Deibel}, A., {et~al.} 2016, ApJ, 831, 13

\bibitem[{{Feroci} {et~al.}(2016){Feroci}, {Bozzo}, {Brandt}, {Hernanz}, {van
  der Klis}, {Liu}, {Orleanski}, {Pohl}, {Santangelo}, {Schanne}, \&
  et~al.}]{feroci2016}
{Feroci}, M., {et~al.} 2016, in Proc. SPIE, Vol. 9905, , 99051R

\bibitem[{{Fujimoto} {et~al.}(1981){Fujimoto}, {Hanawa}, \&
  {Miyaji}}]{fujimoto1981}
{Fujimoto}, M.~Y., {et~al.} 1981, ApJ, 247, 267

\bibitem[{{Galloway} {et~al.}(2008){Galloway}, {Muno}, {Hartman}, {Psaltis}, \&
  {Chakrabarty}}]{galloway2008}
{Galloway}, D.~K., {et~al.} 2008, ApJS, 179, 360

\bibitem[{{Gendreau} {et~al.}(2012){Gendreau}, {Arzoumanian}, \&
  {Okajima}}]{gendreau2012}
{Gendreau}, K.~C., {et~al.} 2012, in Proc. SPIE, Vol. 8443, , 844313

\bibitem[{{Gottwald} {et~al.}(1986){Gottwald}, {Haberl}, {Parmar}, \&
  {White}}]{gottwald1986}
{Gottwald}, M., {et~al.} 1986, ApJ, 308, 213

\bibitem[{Grindlay {et~al.}(1976)Grindlay, Gursky, Schnopper, Parsignault,
  Heise, Brinkman, \& Schrijver}]{grindlay1976}
Grindlay, J., {et~al.} 1976, ApJ, 205, L127

\bibitem[{{Hansen} \& {van Horn}(1975)}]{hansen1975}
{Hansen}, C.~J., Van Horn, H.~M. 1975, ApJ, 195, 735

\bibitem[{{in 't Zand} {et~al.}(2011){in 't Zand}, {Serino}, {Kawai}, \&
  {Heinke}}]{zand11}
{in 't Zand}, J., {et~al.} 2011, ATel, 3625

\bibitem[{{in 't Zand} {et~al.}(2003){in 't Zand}, {Kuulkers}, {Verbunt},
  {Heise}, \& {Cornelisse}}]{zand03}
{in 't Zand}, J.~J.~M., {et~al.} 2003, A\&A, 411, L487

\bibitem[{{in 't Zand} {et~al.}(2004){in 't Zand}, {Cornelisse}, \&
  {Cumming}}]{zand2004}
{in 't Zand}, J.~J.~M., {et~al.} 2004, A\&A, 426, 257

\bibitem[{{in 't Zand} {et~al.}(2005){in 't Zand}, {Cumming}, {van der Sluys},
  {Verbunt}, \& {Pols}}]{zand05}
{in 't Zand}, J.~J.~M., {et~al.} 2005, A\&A, 441, 675

\bibitem[{{in 't Zand} {et~al.}(2007){in 't Zand}, {Jonker}, \&
  {Markwardt}}]{zand2007}
{in 't Zand}, J.~J.~M., {et~al.} 2007, A\&A, 465, 953

\bibitem[{{in 't Zand} \& {Weinberg}(2010)}]{zand10}
{in 't Zand}, J.~J.~M., Weinberg, N.~N. 2010, A\&A, 520, A81

\bibitem[{{in 't Zand} {et~al.}(2014){in 't Zand}, {Cumming}, {Triemstra},
  {Mateijsen}, \& {Bagnoli}}]{zand2014a}
{in 't Zand}, J.~J.~M., {et~al.} 2014, A\&A, 562, A16

\bibitem[{{in 't Zand} {et~al.}(2015){in 't Zand}, {Altamirano}, {Ballantyne},
  {Bhattacharyya}, {Brown}, {Cavecchi}, {Chakrabarty}, {Chenevez}, {Cumming},
  {Degenaar}, {Falanga}, {Galloway}, {Heger}, {Jos{\'e}}, {Keek}, {Linares},
  {Mahmoodifar}, {Malone}, {M{\'e}ndez}, {Miller}, {Paerels}, {Poutanen},
  {R{\'o}zanska}, {Schatz}, {Serino}, {Strohmayer}, {Suleimanov}, {Thielemann},
  {Watts}, {Weinberg}, {Woosley}, {Yu}, {Zhang}, \& {Zingale}}]{zand2015}
{in 't Zand}, J.~J.~M., {et~al.} 2015, ArXiv:1501.02776

\bibitem[{{Iwakiri} {et~al.}(2016a){Iwakiri}, {in 't Zand}, \&
  {Serino}}]{iwakiri2016a}
{Iwakiri}, W., {et~al.} 2016a, these proceedings

\bibitem[{{Iwakiri} {et~al.}(2016b){Iwakiri}, {in 't Zand}, \&
  {Serino}}]{iwakiri2016b}
{Iwakiri}, W., {et~al.} 2016b, ATel, 9882

\bibitem[{{Jager} {et~al.}(1997){Jager}, {Mels}, {Brinkman}, {Galama},
  {Goulooze}, {Heise}, {Lowes}, {Muller}, {Naber}, {Rook}, {Schuurhof},
  {Schuurmans}, \& {Wiersma}}]{jager1997}
{Jager}, R., {et~al.} 1997, A\&As, 125, 557

\bibitem[{{Keek}(2012)}]{keek2012b}
{Keek}, L. 2012, ApJ, 756, 130

\bibitem[{{Keek} \& {Heger}(2011)}]{keek2011}
{Keek}, L., Heger, A. 2011, ApJ, 743, 189

\bibitem[{{Keek} \& {Heger}(2016)}]{keek2016}
{Keek}, L., Heger, A. 2016, MNRAS, 456, L11

\bibitem[{{Keek} {et~al.}(2006){Keek}, {in't Zand}, \& {Cumming}}]{keek2006}
{Keek}, L., {et~al.} 2006, A\&A, 455, 1031

\bibitem[{{Keek} {et~al.}(2008){Keek}, {in 't Zand}, {Kuulkers}, {Cumming},
  {Brown}, \& {Suzuki}}]{keek2008}
{Keek}, L., {et~al.} 2008, A\&A, 479, 177

\bibitem[{{Keek} {et~al.}(2009){Keek}, {Langer}, \& {in 't Zand}}]{keek2009}
{Keek}, L., {et~al.} 2009, A\&A, 502, 871

\bibitem[{{Keek} {et~al.}(2012){Keek}, {Heger}, \& {in 't Zand}}]{kee12a}
{Keek}, L., {et~al.} 2012, ApJ, 752, 150

\bibitem[{{Keek} {et~al.}(2014){Keek}, {Ballantyne}, {Kuulkers}, \&
  {Strohmayer}}]{keek2014}
{Keek}, L., {et~al.} 2014, ApJL, 797, L23

\bibitem[{{Keek} {et~al.}(2015){Keek}, {Cumming}, {Wolf}, {Ballantyne},
  {Suleimanov}, {Kuulkers}, \& {Strohmayer}}]{keek2015}
{Keek}, L., {et~al.} 2015, MNRAS, 454, 3559

\bibitem[{{Keek} {et~al.}(2016){Keek}, {Wolf}, {Ballantyne}}]{keek2016b}
{Keek}, L., {et~al.} 2016, ApJ, 826, 79

\bibitem[{{Kuulkers}(2002)}]{kuu02c}
{Kuulkers}, E. 2002, A\&A, 383, L5

\bibitem[{{Kuulkers}(2009)}]{kuu09b}
{Kuulkers}, E. 2009, ATel, 2140

\bibitem[{{Kuulkers} {et~al.}(2002){Kuulkers}, {in 't Zand}, {van Kerkwijk},
  {Cornelisse}, {Smith}, {Heise}, {Bazzano}, {Cocchi}, {Natalucci}, \&
  {Ubertini}}]{kuu02a}
{Kuulkers}, E., {et~al.} 2002, A\&A, 382, 503

\bibitem[{{Kuulkers} {et~al.}(2004){Kuulkers}, {in't Zand}, {Homan}, {van
  Straaten}, {Altamirano}, \& {van der Klis}}]{kuulkers2004}
{Kuulkers}, E., {et~al.} 2004, in AIP Series, Vol. 714, X-ray Timing 2003:
  Rossi and Beyond, ed. P.~{Kaaret}, F.~K. {Lamb}, \& J.~H. {Swank}, 257--260

\bibitem[{{Kuulkers} {et~al.}(2010){Kuulkers}, {in 't Zand}, {Atteia},
  {Levine}, {Brandt}, {Smith}, {Linares}, {Falanga},
  {S{\'a}nchez-Fern{\'a}ndez}, {Markwardt}, {Strohmayer}, {Cumming}, \&
  {Suzuki}}]{kuu10}
{Kuulkers}, E., {et~al.} 2010, A\&A, 514, A65

\bibitem[{{Lamb} \& {Lamb}(1978)}]{lamblamb1978}
{Lamb}, D.~Q., Lamb, F.~K. 1978, ApJ, 220, 291

\bibitem[{{Lewin} {et~al.}(1984){Lewin}, {Vacca}, \& {Basinska}}]{lewin1984}
{Lewin}, W.~H.~G., {et~al.} 1984, ApJL, 277, L57

\bibitem[{{Lewin} {et~al.}(1993){Lewin}, {van Paradijs}, \& {Taam}}]{lewin1993}
{Lewin}, W.~H.~G., {et~al.} 1993, SSR, 62, 223

\bibitem[{{Makishima} {et~al.}(1981){Makishima}, {Ohashi}, {Inoue}, {Koyama},
  {Matsuoka}, {Murakami}, {Oda}, {Ogawara}, {Shibazaki}, {Tanaka}, {Kondo},
  {Hayakawa}, {Kunieda}, {Makino}, {Masai}, {Nagase}, {Tawara}, {Miyamoto},
  {Tsunemi}, \& {da Yamashita}}]{makishima1981}
{Makishima}, K., {et~al.} 1981, ApJL, 247, L23

\bibitem[{{Matsuoka} {et~al.}(1980){Matsuoka}, {Inoue}, {Koyama}, {Makishima},
  {Murakami}, {Oda}, {Ogawara}, {Ohashi}, {Shibazaki}, {Tanaka}, {Kondo},
  {Hayakawa}, {Kunieda}, {Makino}, {Masai}, {Nagase}, {Tawara}, {Miyamoto},
  {Tsunemi}, \& {Yamashita}}]{matsuoka1980}
{Matsuoka}, M., {et~al.} 1980, ApJL, 240, L137

\bibitem[{{Medin} \& {Cumming}(2011)}]{medin2011}
{Medin}, Z., Cumming, A. 2011, ApJ, 730, 97

\bibitem[{{Motch} {et~al.}(1987)}]{motch1987}
Motch, C., et al. 1987, ApJ, 313, 792

\bibitem[{{Nelson} {et~al.}(1986){Nelson}, {Rappaport}, \& {Joss}}]{nrj86}
{Nelson}, L.~A., {et~al.} 1986, ApJ, 304, 231

\bibitem[{{Peng} {et~al.}(2007){Peng}, {Brown}, \& {Truran}}]{peng2007}
{Peng}, F., {et~al.} 2007, ApJ, 654, 1022

\bibitem[{{Priedhorsky} {et~al.}(1986)}]{pried1986}
Priedhorsky, W., et al. 1986, IAUC, 4247

\bibitem[{{Bandyopadhyay} {et~al.}(2002)}]{reba2002}
Bandyopadhyay, R.M., et al. 2002, ApJ, 570, 793

\bibitem[{{Schatz} {et~al.}(2003){Schatz}, {Bildsten}, \&
  {Cumming}}]{schatz2003}
{Schatz}, H., {et~al.} 2003, ApJL, 583, L87

\bibitem[{{Schatz} {et~al.}(2014){Schatz}, {Gupta}, {M{\"o}ller}, {Beard},
  {Brown}, {Deibel}, {Gasques}, {Hix}, {Keek}, {Lau}, {Steiner}, \&
  {Wiescher}}]{schatz2014}
{Schatz}, H., {et~al.} 2014, Nat, 505, 62

\bibitem[{{Serino} {et~al.}(2012){Serino}, {Mihara}, {Matsuoka}, {Nakahira},
  {Sugizaki}, {Ueda}, {Kawai}, \& {Ueno}}]{serino2012}
{Serino}, M., {et~al.} 2012, PASJ, 64, 91

\bibitem[{{Serino} {et~al.}(2016){Serino}, {Iwakiri}, {Tamagawa}, {Sakamoto},
  {Nakahira}, {Matsuoka}, {Yamaoka}, \& {Negoro}}]{serino2016}
{Serino}, M., {et~al.} 2016, PASJ, 68, 95

\bibitem[{{Shahbaz} {et~al.}(2008)}]{shabaz2008}
Shabaz, T., et al. 2008, PASP, 120, 848

\bibitem[{{Stevens} {et~al.}(2014){Stevens}, {Brown}, {Cumming}, {Cyburt}, \&
  {Schatz}}]{stevens2014}
{Stevens}, J., {et~al.} 2014, ApJ, 791, 106

\bibitem[{{Strohmayer} \& {Bildsten}(2006)}]{stroh2006}
{Strohmayer}, T., Bildsten, L. 2006, {New views of thermonuclear bursts} (Compact
  stellar X-ray sources), 113--156

\bibitem[{{Strohmayer} \& {Brown}(2002)}]{stroh02a}
{Strohmayer}, T.~E., Brown, E. 2002, ApJ, 566, 1045

\bibitem[{{Strohmayer} \& {Markwardt}(2002)}]{stroh02b}
{Strohmayer}, T.~E., Markwardt, C.~B. 2002, ApJ, 577, 337

\bibitem[{{Strohmayer} \& {Mahmoodifar}(2014)}]{stroh14}
{Strohmayer}, T.~E., Mahmoodifar, S. 2014, ApJL, 793, L38

\bibitem[{{van Paradijs} {et~al.}(1990)}]{jvp90}
Van Paradijs, J., et al. 1990, A\&A, 234, 181

\bibitem[{{Wallace} \& {Woosley}(1981)}]{wallace1981}
{Wallace}, R.~K., Woosley, S.~E. 1981, ApJS, 45, 389

\bibitem[{{Weinberg} \& {Bildsten}(2007)}]{weinberg2007}
{Weinberg}, N.~N., Bildsten, L. 2007, ApJ, 670, 1291

\bibitem[{{Weinberg} {et~al.}(2006){Weinberg}, {Bildsten}, \&
  {Brown}}]{weinberg2006b}
{Weinberg}, N.~N., {et~al.} 2006, ApJL, 650, L119

\bibitem[{{Wijnands}(2001)}]{wijnands2001}
{Wijnands}, R. 2001, ApJL, 554, L59

\bibitem[{{Wilson-Hodge} {et~al.} (2017){Wilson-Hodge}, C.~A. and
    {Ray}, P.~S. and {Gendreau}, K. and {Chakrabarty}, D. and
    {Feroci}, M. and {Maccarone}, T. and {Arzoumanian}, Z. and
    {Remillard}, R.~A. and {Wood}, K. and {Griffith}, C. and {STROBE-X
      Collaboration}}]{wilson2017}
{Wilson-Hodge}, C.~A., et al. 2017, AAS Meeting Abstracts, 229:339.04

\bibitem[{{Zhang} {et~al.}(2016){Zhang}, {Feroci}, {Santangelo}, {Dong},
  {Feng}, {Lu}, {Nandra}, {Wang}, {Zhang}, {Bozzo}, {Brandt}, {De Rosa}, {Gou},
  {Hernanz}, {van der Klis}, {Li}, {Liu}, {Orleanski}, {Pareschi}, {Pohl},
  {Poutanen}, {Qu}, {Schanne}, {Stella}, {Uttley}, \& {Watts}}]{zhang2016}
{Zhang}, S.~N., {et~al.} 2016, in Proc. SPIE, Vol. 9905, , 99051Q

\end{thebibliography}
\end{document}